\documentclass{ptephy_v1}
\usepackage{tikz,tikz-3dplot}
\tdplotsetmaincoords{80}{45}

\tikzset{surface/.style={draw=black, fill=white, fill opacity=.6}}

\usepackage{tikz}
\usetikzlibrary{decorations.pathmorphing,patterns}

\usepackage{amssymb}
\usepackage{amsmath,amscd}
\usepackage{color}
\usepackage{mathrsfs}

\usepackage{bm}

\usepackage[normalem]{ulem}

\begin{document}

\title{AdS/BCFT correspondence and Lovelock theory in the presence of canonical scalar field}
\author{Oleksii Sokoliuk$^{1,2}$}
\author{Fabiano F. Santos$^{3}$}
\author{Alexander Baransky$^{2}$}

\affil{%
$^1$ Main Astronomical Observatory of the NAS of Ukraine (MAO NASU),
Kyiv, 03143, Ukraine\\
$^2$ Astronomical Observatory, Taras Shevchenko National University of Kyiv,
3 Observatorna St., 04053 Kyiv, Ukraine\\
$^3$ Institute of Theoretical Physics, Beijing University of Technology, Beijing 100124, ChinaInstituto de Física,Universidade Federal do Rio de Janeiro, Caixa Postal 68528, Rio de Janeiro-RJ, 21941-972 -- Brazil\\
}


\begin{abstract}
In the current study, authors present the comprehensive analysis of the Anti-de Sitter/Boundary Conformal Field Theory (AdS/BCFT) correspondence for Lovelock theory of second order in the presence of canonical (and real) scalar field non-minimally coupled to modified gravitation sector. Authors consider the special case with three dimensional BTZ (Banados-Teitelboim-Zanelli) black hole and dual 2D Conformal Field Theory with the additional boundary for higher dimensional AdS spacetime. Moreover, for the given case additional boundary profile was investigated, as well as the holographic renormalization procedure for Lovelock-scalar theory. Besides, from holographic renormalization black hole entropy was obtained numerically. Finally, from the entropy and black hole temperature dual to BCFT some thermodynamical quantities, such as heat capacity, sound speed, energy-momentum trace and Hawking-Page phase transition were probed.
\end{abstract}

\maketitle

%
\section{Introduction}\label{sec:1}
It is well-known that Einsteinian General Theory of Relativity (further - GR) generally could describe the universe evolution with high prevision and GR has passed numerous theoretical and observational tests within Solar System, Milky Way and whole observable universe. If one will introduce additional matter fields, such as inflaton or $\Lambda$ term, cosmological inflation and late-time accelerated expansion of the universe scenario could be reconstructed. However, there are some critical problems present in the GR, such as dark matter, dark energy problems (presence of Lambda term leads to various issues \cite{doi:10.1142/S0218271800000542,PADMANABHAN2003235}). In order to properly address the aforementioned problems, during the few past decades method of gravity modification were used. With the such method, new geometrodynamical terms are introduced in the Einstein-Hilbert action within gravitational sector (such as Riemann tensor or function of Ricci curvature \cite{Gottlober_1990,PhysRevD.43.965,Amendola_1993}, Einstein tensor \cite{Horndeski:1974wa,PhysRevLett.108.051101,PhysRevD.85.104040}, torsion \cite{Capozziello:2011et,Hehl:1976kj,Hayashi:1979qx} and non-metricity \cite{BeltranJimenez:2019tme,Amelino-Camelia:2011lvm,BeltranJimenez:2019esp} with unique affine connections). In the current article, we will concentrate our study on the one special kind of modified gravity, namely Lovelock theory, which has gained much interest in the last years, and has been studied in the theoretical \cite{Myers:1988ze,Dong:2013qoa,Cai:2003kt,Camanho:2009hu} and experimental context \cite{PhysRevD.103.064002,doi:10.1142/S0218271818500840}.

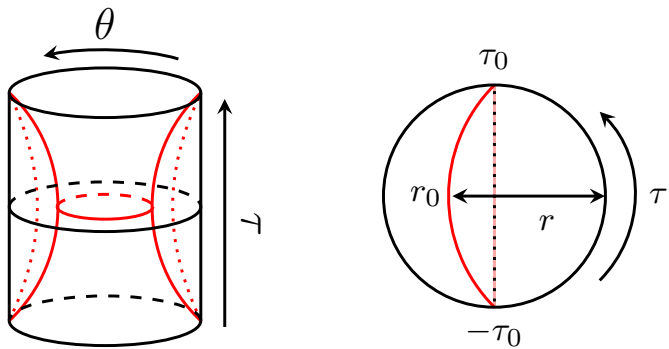
\begin{figure}[!h]
    \centering
\scalebox{1.4}{
\def\a{45}
\tdplotsetmaincoords{75}{\a}
\begin{tikzpicture}
	[scale=7,
	tdplot_main_coords,
	axis/.style={->,black,very thin},
	curve/.style={black,thin}]
	\def\radius{.13}
	\def\axissize{0.3}
	\def\th{0.32}

	\tdplotsinandcos{\sintheta}{\costheta}{\a}
	\draw[thick] (\radius*\costheta,\radius*\sintheta,0) {} -- (\radius*\costheta,\radius*\sintheta,\th);
	\draw[thick,color=red] (\radius*\costheta,\radius*\sintheta,0) {}  to [bend left=47] (\radius*\costheta,\radius*\sintheta,\th);
	\draw[thick,dotted,color=red] (\radius*\costheta,\radius*\sintheta,0) {} to [bend left=25]
	(\radius*\costheta,\radius*\sintheta,\th);
	\draw[thick,-stealth] (1.5*\radius*\costheta,\radius*\sintheta,0) {} to 
	(1.5*\radius*\costheta,\radius*\sintheta,\th);
    \node[rotate=-90] at (-0.22,0.5) 
    {\large $\tau$};
    \node[] at (-1.1,1.1) 
    {\large $\theta$};
	\tdplotsinandcos{\sintheta}{\costheta}{\a+180}
	\draw[thick] (\radius*\costheta,\radius*\sintheta,0) -- (\radius*\costheta,\radius*\sintheta,\th);
	\draw[thick,color=red] (\radius*\costheta,\radius*\sintheta,0) {} to [bend right=47]
	(\radius*\costheta,\radius*\sintheta,\th);
	\draw[thick,dotted,color=red] (\radius*\costheta,\radius*\sintheta,0) {} to [bend right=25]
	(\radius*\costheta,\radius*\sintheta,\th);
	
	\tdplotsinandcos{\sintheta}{\costheta}{0}
	\tdplotdrawarc[curve,thick]{(0,0,\th)}{\radius*\costheta}{\a-360}{\a}{}{}
	\tdplotdrawarc[curve,thick]{(0,0,\th/2)}{\radius*\costheta}{\a-180}{\a}{}{};
	\tdplotdrawarc[curve,thick,dashed]{(0,0,\th/2)}{\radius*\costheta}{\a+180}{\a}{}{}
	\tdplotdrawarc[curve,thick,dashed]{(0,0,0)}{\radius*\costheta}{\a}{\a+180}{}{}
	\tdplotdrawarc[curve,thick]{(0,0,0)}{\radius*\costheta}{\a}{\a-180}{}{}
	\tdplotdrawarc[curve,thick,dashed,color=red]{(0,0,\th/2)}{0.5*\radius*\costheta}{\a+180}{\a}{}{}
	\tdplotdrawarc[curve,thick,color=red]{(0,0,\th/2)}{0.5*\radius*\costheta}{\a-180}{\a}{}{}
	\tdplotdrawarc[curve,thick,stealth-]{(0,0,0.025+\th)}{\radius*\costheta}{\a+130}{\a+40}{}{}
\end{tikzpicture}
\hspace{+2em}

\begin{tikzpicture}[scale=1,
	axis/.style={->,black,very thin},
	curve/.style={black,thin}]
    \draw[thick,fill,red!35,shift={(0 cm,0.65 cm)}] (0,-0.7)--(0,1.4);

	\node[circle, draw=black,thick,
      minimum size=60pt, outer sep=0pt,,opacity=0,shift={(0 cm,1 cm)}](circ){};
\draw[thick,red,opacity=0,shift={(0 cm,1 cm)}]
    (circ.south) to [bend left = 45] (circ.north);
    
    \draw[solid, shift={(0 cm,2 cm)},opacity=0] (2,-1.5) circle (1);
\draw[thick,red,shift={(0 cm,1 cm)}]
    (circ.south) to [bend left = 45] (circ.north);
    	\node[shift={(0 cm,1 cm)},circle, draw=black,thick,
      minimum size=60pt, outer sep=0pt,](circ){};
\draw[thick,-stealth] (0,1) -- (1.05,1);
\node[] at (0.5,0.75) 
    {\small $r$};
\draw[thick,dotted] (0,0) -- (0,2);
\draw[thick,-stealth] (0,1) -- (-0.4,1);
\node[] at (-0.65,1) 
    {\small $r_0$};
\draw[thick,-stealth] (1,0.2) to [bend right=45] (1,1.8);
\node[] at (1.56,1) 
    {\small $\tau$};
\node[] at (0,2.3) 
    {\small $\tau_0$};
\node[] at (0,-0.3) 
    {\small $-\tau_0$};

\end{tikzpicture}
}
    \caption{EoW brane embedded into the higher dimensional AdS spacetime (bulk). Solid line represents braneworlds with positive brane tension, dotted with vanishing tension}
    \label{fig:1}
\end{figure}

Apart from the Lovelock theory, we as well considered Anti-de Sitter/Conformal Field Theory correspondence (i.e. duality) in the current study (for the detailed discussion on AdS/CFT, refer to \cite{Maldacena:1997re,Maldacena:1997re,AHARONY2000183}). These works shed light on the problems of theoretical physics and cosmology and made it possible to study strongly interacting systems in more detail. For example, now it is possible to compute the entalegement entropy of conformal boundary region from it's holographic dual via AdS/CFT correspondance \cite{Ryu:2006bv,Ryu:2006ef,Hubeny:2007xt}, holographic complexity of the spacetime \cite{Brown:2015bva,Alishahiha:2015rta,Carmi:2016wjl} and many more important quantities. Scheme for AdS/CFT correspondance is respectively placed on the Figure (\ref{fig:1}), where EoW brane denotes the conformal boundary, $\tau$ - conformal time (cyclic variable).

\begin{figure}[!h]
    \centering
    \scalebox{1.4}{
    \pgfkeys{/tikz/.cd, view angle/.initial=0, view angle/.store in=\picangle}
\tikzset{
  horizontal/.style={y={(0,sin(\picangle))}},
  vertical at/.style={x={([horizontal] #1:1)}, y={(0,cos(\picangle)cm)}},
  every label/.style={font=\tiny, inner sep=1pt},
  shorten/.style={shorten <=#1, shorten >=#1},
  shorten/.default=3pt,
  ->-/.style={decoration={markings, mark=at position #1 with {\arrow{>}}}, postaction={decorate}},
  ->-/.default=0.5
}

\begin{tikzpicture}[scale=1.34, view angle=15, >=stealth]
  \draw[fill=red!50, fill opacity=.2, horizontal, thick] (0,0) circle (1);
 (0:-105:1);
  \draw[fill=red, fill opacity=.3, color=red, thick] (0:1) arc (0:-180:1) [horizontal] arc (-180:0:1);
  \draw[fill=gray!50, fill opacity=.0, horizontal, color = black, thick] (0,0) circle (1);
    \draw[fill=red, fill opacity=.0, color=black, thick] (0:1) arc (0:-180:1) [horizontal] arc (-180:0:1);
 (0:-105:1);

    \node[] at (0,0) 
    { $\mathrm{CFT}$};

    \node[] at (0,-0.7) 
    {$\mathcal{M}_{\mathrm{AdS}}$};
    \node[] at (1.1,-0.55) 
    {$Q$};
\end{tikzpicture}
}
    \caption{AdS/CFT correspondance in the presence of boundary hypersurface $Q$}
    \label{fig:2}
\end{figure}
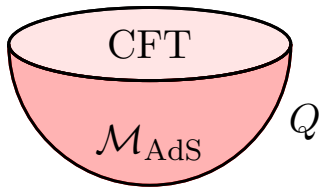

On the other hand, the inclusion of the additional boundary for higher dimensional AdS spacetime could lead to the interesting results with the use of the so-called AdS/BCFT correspondence (where latter B stands for Boundary). AdS/BCFT formalism were firstly presented in the pioneering works of \cite{PhysRevLett.107.101602,Fujita:2011fp}. Setup, that were used in the present study to probe AdS/BCFT is consequently introduced schematically on the Figure (\ref{fig:2}). As we see, apart from the CFT that lives on the boundary $r=0$, we have additional boundary of $\mathcal{M}_{\mathrm{AdS}}$, namely $Q$ (which is not necessary asymptotically Anti-de Sitter). In the following section, we are going to present the methodology that were used.
\section{Methodology}\label{sec:2}
Motivated by the recent studies and applications of the AdS/CFT duality in the Einstein gravity and beyond, as for example in Horndeski gravity, we propose some applications of this correspondence in Lovelock theory in the presence of canonical scalar field. In our scenario we consider the BTZ black hole as the theoretical background for computation of all thermodynamical quantities. In this way, we present the organization of the quantities the will be analyzed in the further investigation:
\begin{itemize}
    \item First we present the Lovelock theory in the presence of canonical scalar field and in sequence we show the form Gibbons-Hawking-York surface term for the Lovelock gravitation. Furthermore, solving numerically the equation of motion for the BCFT$_{2}$, we found the boundary profile for the BTZ black hole geometry;
    \item Trough the holographic renormalization scheme that we applied, the free energy in the Lovelock theory in the presence of canonical scalar field for the AdS BTZ black hole has been computed properly;
    \item From this free energy, we derived various thermodynamic quantities, such as: the entropy of the BTZ black hole that has contributions of the boundary, heat capacity, speed of sound and trace of the energy-momentum tensor. During the analysis of these quantities we discuss the influence of the Lovelock coupling parameter;
    \item Finally, as the last quantity that comes from the free energy we computed Hawking-Page phase transition and probed its behavior.
\end{itemize}

\subsection{Article organisation}
This article is organised as follows: in the first Section (\ref{sec:1}) we introduce the problems of Einsteinian gravitation and the possible ways to solve them, present the formalism of AdS/CFT and AdS/BCFT. In the Section (\ref{sec:2}) we mention the methodology, that were used to numerically investigate the model. In the third Section we present the formalism of Lovelock extended General Theory of Relativity non-minimally coupled to the canonical scalar field. On the other hand, AdS$_3$/BCFT$_2$ correspondence is studied in the Section (\ref{sec:4}) for Lovelock theory of gravitation. Besides, in the next Section we use results from (\ref{sec:4}) to probe the three-dimensional asymptotically AdS BTZ black hole in the sense of AdS/BCFT. Finally, in the Sections (\ref{sec:6}) and (\ref{sec:7}) we use the method of holographic renormalization of Lovelock theory and with the help of such procedure we derive various thermodynamical quantities of our consideration. Concluding remarks on the key topics of our study are consequently provided in the Section (\ref{sec:8}). 

\section{Lovelock extension of GR}\label{sec:3}
In the present paper as a background geometry we assume the so-called modified Lovelock gravity, for which Einstein-Hilbert action integral is given by \cite{DeFelice:2010aj}
\begin{equation}
    \mathcal{S}[g,\Gamma,\Psi_i]=\int d^4x\sqrt{-g}\bigg[\kappa R-2g^{\mu\nu}\partial_\mu\phi\partial_\nu\phi+\alpha f(\phi)\mathcal{L}_{\mathrm{GB}}\bigg]+\mathcal{S}_\mathrm{M}[g,\Psi_i]
    \label{eq:2.1}
\end{equation}
Where $\Gamma$ is the well-known torsion free and metric compatible Levi-Cevita affine connection, $g=\prod^3_{\mu,\nu=0}g_{\mu\nu}=\det g_{\mu\nu}$ is the metric tensor determinant, $R$ - Ricci scalar curvature made from connection, $\phi$ is scalar field non-minimally coupled to the gravity through the coupling function $f(\phi)$, $\mathcal{L}_{\mathrm{GB}}$ is the contribution of Lovelock gravity (up to the second order, more precisely we consider Gauss-Bonnet case):
\begin{equation}
    \mathcal{L}_{\mathrm{GB}}=R^2-4R_{\mu\nu}R^{\mu\nu}+R_{\mu\nu\alpha\beta}R^{\mu\nu\alpha\beta}
\end{equation}
As well, in the equations above we define $\alpha$ as a additional degree of freedom, that defines the contribution of Lovelock terms to the total action, $\mathcal{S}_\mathrm{M}[g,\Psi_i]$ is the action integral for additional matter fields $\Psi_i$ that minimally, non-minimally coupled to gravitation sector for non-vacuum case. By varying action integral (\ref{eq:2.1}) with respect to the metric tensor inverse $g^{\mu\nu}$ one could obtain effective field equations for our particular theory of gravitation:
\begin{equation}
        G_{\mu\nu} + \Gamma_{ \mu \nu } = 2 \nabla_{ \mu } \phi \nabla_{ \nu } \phi - g_{ \mu \nu } \nabla_{ \alpha } \phi \nabla^{ \alpha } \phi +T_{\mu\nu}
\end{equation}
Where \cite{PhysRevD.101.044054,Danchev:2021tew}
\begin{equation}
    G_{\mu\nu}=R_{\mu\nu}-\frac{1}{2}Rg_{\mu\nu}
\end{equation}
\begin{equation}
    \begin{gathered}
        \Gamma_{ \mu \nu } = - R( \nabla_{ \mu } \Psi_{ \nu } + \nabla_{ \nu } \Psi_{ \mu } ) - 4 \nabla^{ \alpha } \Psi_{ \alpha } \left( R_{ \mu \nu } - \frac{ 1 }{ 2 } R g_{ \mu \nu } \right) + 4 R_{ \mu \alpha } \nabla^{ \alpha } \Psi_{ \nu } + 4 R_{ \nu \alpha } \nabla^{ \alpha } \Psi_{ \mu }  \\
    - 4 g_{ \mu \nu } R^{ \alpha \beta } \nabla_{ \alpha } \Psi_{ \beta } + 4 R^{ \beta }_{ \mu \alpha \nu } \nabla^{ \alpha } \Psi_{ \beta }
    \end{gathered}
\end{equation}
\begin{equation}
    \Psi_{ \mu } = \alpha \frac{ df( \phi ) }{ d \phi } \nabla_{ \mu } \phi
\end{equation}
As well, $T_{\mu\nu}$ is energy-momentum tensor, which describes the contribution of matter to the field equations and generally defined as follows:
\begin{equation}
    T_{\mu\nu}=-\frac{2}{\sqrt{-g}}\frac{\delta (\sqrt{-g}\mathcal{L}_\mathrm{M})}{\delta g^{\mu\nu}}
\end{equation}
Since we already defined all of the necessary quantities of our background model, we could move further and investigate AdS$_3$ spacetimes on the CFT$_2$ with the present boundary.
\section{AdS$_3$/BCFT$_2$ in the Lovelock theory}\label{sec:4}
In the current section we are going to study the well-known AdS/CFT correspondence in the presence of boundary. Contribution to the total action of the manifold boundary is called Gibbons-Hawking-York surface term \cite{PhysRevLett.107.101602,Fujita:2011fp}. Such term in the general form of Lovelock theory looks exactly like \cite{PhysRevD.101.124045}:
\begin{equation}
    \mathcal{S}_{\mathrm{GHY}}=\int d^{2}x\sqrt{-h}\bigg(K+\alpha f(\phi)\delta^{i_1i_2i_3}_{j_1j_2j_3}K^{j_1}_{i_1}\bigg(\overline{R}^{j_2j_3}_{\:\:\:\:\:\:\:\:i_2i_3}-\frac{2\epsilon}{3}K^{j_2}_{i_2}K^{j_3}_{i_3}\bigg)+\mathcal{L}_\mathrm{M}\bigg)
    \label{eq:3.1}
\end{equation}
Where $K=h^{ab}K_{ab}$ is extrinsic curvature of the manifold where $K_{ab}=e^\mu_a e^\nu_b\nabla_\mu n_\nu$ and $h_{ab}$ is the induced on the boundary metric tensor, $n_a$ - normal vector to the boundary hypersurface, $\delta^{\alpha_1,\alpha_2,...,\alpha_n}_{\beta_1,\beta_2,...,\beta_n}=n!\delta^{\alpha_1}_{[\beta_1}...\delta^{\alpha_n}_{\beta_n]}$ - generalized Kronecker symbol. Alternatively, boundary action could also be expressed as:
\begin{equation}
     \mathcal{S}_{\mathrm{GHY}}=\int d^{2}x\sqrt{-h}\bigg(K+2\alpha f(\phi)(J-2\overline{G}_{ij}K^{ij})+\mathcal{L}_\mathrm{M}\bigg)
\end{equation}
Where $J$ is the trace of tensor $J_{ij}$:
\begin{equation}
    \epsilon J_{ij}=-\frac{2}{3}K_{il}K^{lp}K_{pj}+\frac{2}{3}KK_{il}K^l_{j}+\frac{1}{3}K_{ij}(K^{lp}K_{lp}-K^2)
\end{equation}
Induced metric consequently could be derived from the projecting tensor:
\begin{equation}
    h_{\mu\nu}=g_{\mu\nu}-\epsilon n_\mu n_\nu,\quad h_{ab}=e^\mu_a e^\nu_b h_{\mu\nu},\quad e^\mu_a=\frac{\partial x^\mu}{\partial y^a}
\end{equation}
Where $\epsilon=\pm1$ depending on whether normal is timelike or spacelike. 
Imposing Dirichlet boundary conditions for CFT we could vary GHY term and obtain corresponding field equation for induced metric with the use of Neumann boundary condition and timelike boundary hypersurface (we consider purely quadratic form of $f(\phi)=\alpha/8\phi^{2}$):
\begin{equation}
\begin{gathered}
\frac{1}{12} \Bigl(2 \alpha^2 K_{a}{}^{c} (-3 K_{b}{}^{d} K_{cd} 
+ 2 K_{bc} K^{d}{}_{d}) \phi^2 + 2 K_{ab} (6 + \alpha^2 K_{cd} 
K^{cd} \phi^2) + h_{ab} \bigl(\alpha^2 K_{c}{}^{e} K^{cd} K_{de} 
\phi^2 \\
-  K^{c}{}_{c} (6 + \alpha^2 K_{de} K^{de} 
\phi^2)\bigr)\Bigr)=0
\end{gathered}
\label{eq:3.5}
\end{equation}
Where bar quantities are intrinsic ones built from the induced metric on the boundary $h_{ij}$. As well, we assume that stress-energy-momentum tensor $T_{ab}$ as well as Einstein tensor $G_{ab}$ on the boundary vanish, so solution is vacuum. Finally, we could also obtain field equations for the bulk and boundary actions without any matter fields apart from scalar one present (varying w.r.t. $g^{\mu\nu}$ and $\phi$ respectively):
\begin{equation}
    \begin{gathered}
    \mathcal{A}_{\mu\nu}[g_{\mu\nu},\phi]= G_{\mu\nu} + \Gamma_{ \mu \nu } -2 \nabla_{ \mu } \phi \nabla_{ \nu } \phi + g_{ \mu \nu } \nabla_{ \alpha } \phi \nabla^{ \alpha } \phi 
    \end{gathered}
\end{equation}
\begin{equation}
    \begin{gathered}
    \mathcal{B}[g_{\mu\nu},\phi]=-\nabla_\alpha\nabla^\alpha\phi-\frac{\alpha}{4}\frac{df(\phi)}{d\phi}\mathcal{L}_{\mathrm{GB}}
    \end{gathered}
\end{equation}
\begin{equation}
    \begin{gathered}
    \mathcal{C}[h_{ab},\phi]=\frac{1}{3} \alpha^2 K_{bc} (- K_{a}{}^{c} K^{ab} + K^{a}{}_{a} 
K^{bc}) \phi 
    \end{gathered}
\end{equation}
Remarkably, from the Euler-Lagrange formulation, $\mathcal{B}[g_{\mu\nu},\phi]=\mathcal{C}[g_{ij},\phi]$.
Therefore, while we already derived field equation for induced metric on the boundary hypersurface, we could proceed the the investigation of such hypersurfaces in the background of BTZ black hole.
\section{BTZ black hole as a probe of AdS/BCFT}\label{sec:5}
In the present study, as a probe of Anti-de Sitter/Boundary CFT correspondence among other choices we have chosen the $(2+1)$ dimensional BTZ black hole. For that case, exact form of metric tensor line element is given by the expression below \cite{PhysRevLett.69.1849,PhysRevD.48.1506}:
\begin{equation}
    ds^2 = \frac{L^2}{r^2}(-f(r)dt^2+\frac{dr^2}{f(r)}+dy^2)
\end{equation}
Where $f(r)$ is the BTZ black hole metric function with radial dependence and $L$ is usual AdS radius, that defines the contribution of negative cosmological constant to the spacetime curvature. 
Tetrad field consequently reads:
\begin{equation}
    e^\mu_a=\bigg(\frac{L\sqrt{f(r)}}{r},\frac{L}{r\sqrt{f(r)}},\frac{L}{r}\bigg)
\end{equation}

Assuming that $\phi=\phi(r)$ and imposing $\mathcal{A}_{rr}[g_{rr},\phi]=\mathcal{B}[g_{rr},\phi]=0$ we could derive $\phi'(r)$ for our BTZ black hole solution (we assume that second and higher order derivatives of $f(r)$ and scalar field $\phi$ vanish).

\begin{figure}[!h]
    \centering
    \includegraphics[width=\textwidth]{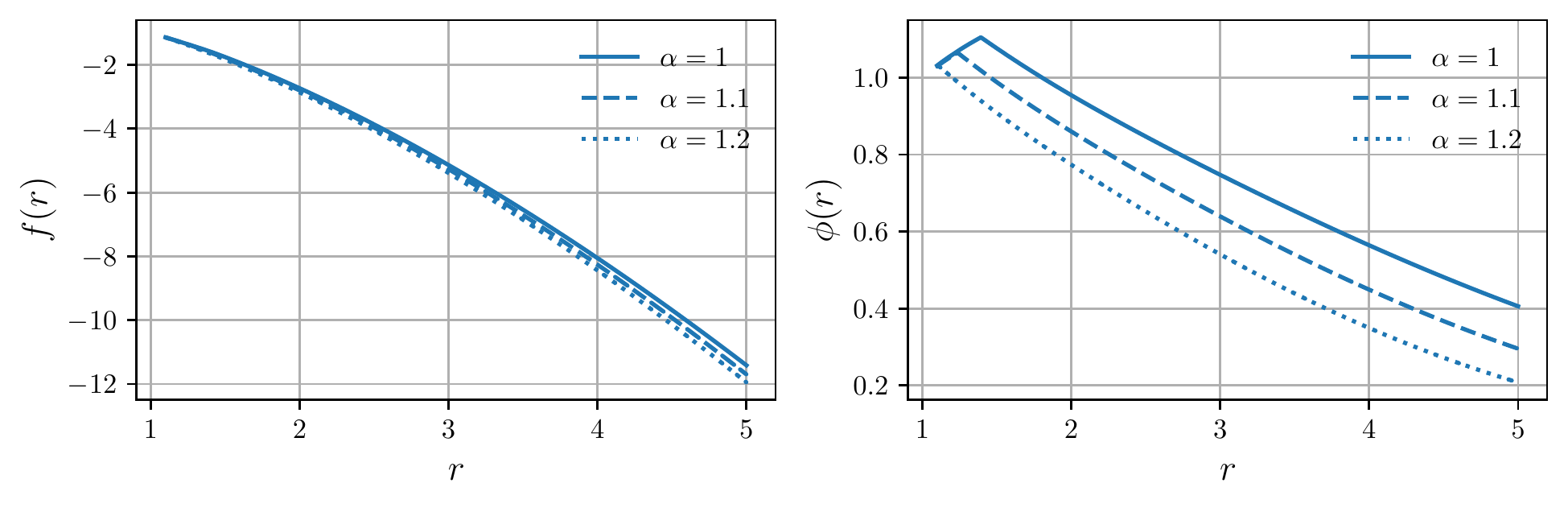}
    \caption{Numerically derived metric function and scalar field for asymptotically AdS$_3$ BTZ black hole with $L=1$ and varying Lovelock coupling $\alpha$}
    \label{fig:3}
\end{figure}

As well, BTZ metric function could be derived numerically imposing $f(r_\mathrm{h})=f(1)<0$. We plot such solutions for unitary AdS radius and different values of Lovelock coupling $\alpha$ on the Figure (\ref{fig:3}). As one could see, metric function behaves as expected for BTZ black holes that are asymptotically AdS. As well, scalar field asymptotically vanish.

For BTZ black holes, Hawking temperature is given in the terms of metric function at the horizon:
\begin{equation}
    T_{\mathrm{H}}=T_{\mathrm{BCFT}}=\frac{1}{4\pi}|f'(r_\mathrm{h})|\approx \frac{1}{2\pi r_\mathrm{h}}
\end{equation}
As we see, Hawking temperature is dual to the BCFT one. 
\subsection{Induced metric}
Following the procedure, described in the article \cite{PhysRevD.104.066014}, we are going to derive the profile of boundary hypersurface with the use of induced metric for asymptotically AdS BTZ black hole:
\begin{equation}
    ds^2_{\mathrm{ind}}=\frac{L^2}{r^2}\bigg(-f(r)dt^2+\frac{g^2(r)dr^2}{f(r)}\bigg)
\end{equation}
Where $g^2(r)=1+y'^2(r)f(r)$ and as usual $y'(r)=dy/dr$. Using aforementioned induced line element, one could as well derive normal vector:
\begin{equation}
    n^\mu = \frac{r}{g(r)L}(0,1,-f(r)y'(r))
\end{equation}
With the use of no-hair theorem (scalar hair must vanish, and therefore $\mathcal{C}[h_{rr},\phi]=0$ holds) and field equation (\ref{eq:3.5}) we could derive the expression for $y'(r)$ and $\phi(r)$ (again, to derive the solution we ignore terms $y''(r)$ and higher).
By solving ODE for $y'(r)$ we could obtain numerical solution for $y(r)$ imposing initial condition $y(r_\mathrm{h})=y_0$.
\begin{figure}[!h]
    \centering
    \includegraphics[width=0.7\textwidth]{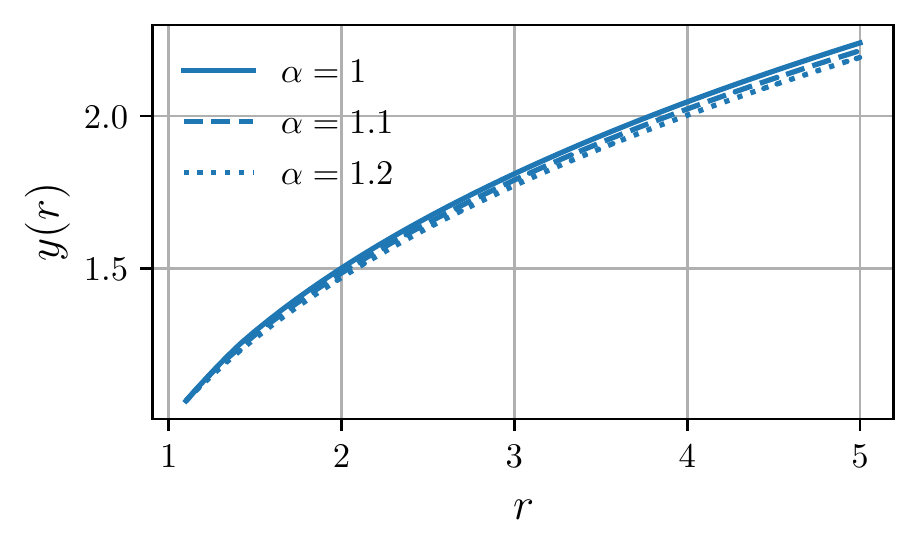}
    \caption{Numerically derived $y(r)$ function for BTZ black hole w.r.t. free parameter $\alpha$ (Gauss-Bonnet coupling constant)}
    \label{fig:4}
\end{figure}

Numerical solution for $y(r)$ is placed on the Figure (\ref{fig:4}). As we see, this function has expected behavior and converge with the results, obtained for Horndeski gravity \cite{PhysRevD.104.066014}. In CFT$_{2}$ is common to refer to $\Delta y=y-y_0$ as the width of the boundary \cite{PhysRevD.104.066014}

\section{Holographic renormalization of Lovelock theory}\label{sec:6}
Now, we will present the holographic renormalization scheme to the Lovelock theory in the AdS/BCFT correspondence. As the first step, we need to compute the euclidean on-shell action for this scenario we need of the boundary counter term to render the finite action, which provide the free energy of thermodynamic systems.

\begin{equation}
     \begin{gathered}
    I_{bulk}=-\frac{1}{16\pi G_{N}}\int_{\mathcal{M}}{d^3x\sqrt{g}\bigg[\kappa R-2g^{\mu\nu}\partial_\mu\phi\partial_\nu\phi+\alpha f(\phi)\mathcal{L}_{\mathrm{GB}}\bigg]}\\
    -\frac{1}{8\pi G_{N}}\int_{\mathcal{M}^{'}}{d^{2}x\sqrt{\Bar{\gamma}}\bigg(K_{(\Bar{\gamma})}+\alpha f(\phi)\delta^{i_1i_2i_3}_{j_1j_2j_3}K^{j_1}_{i_1\:\:\:\:(\Bar{\gamma})}\bigg(\overline{R}^{j_2j_3}_{\:\:\:\:\:\:\:\:i_2i_3}-\frac{2\epsilon}{3}K^{j_2}_{i_2\:\:\:\:(\Bar{\gamma})}K^{(\Bar{\gamma})}_{j_3i_3\:\:\:\:(\Bar{\gamma})}\bigg)\bigg)}\label{eq:H.1}
    \end{gathered}
\end{equation}

In our setup g is the determinant of the metric $g_{\mu\nu}$ on the bulk $\mathcal{M}$ with a induced metric $\Bar{\gamma}$ on $\mathcal{M}^{'}$, respectively, and the trace of the extrinsic curvature is $K_{(\Bar{\gamma})}$. For the boundary side, we have

\begin{equation}
     \begin{gathered}
    I_{bdry}=-\frac{1}{16\pi G_{N}}\int_{\mathcal{M}}{d^3x\sqrt{g}\bigg[\kappa R-2g^{\mu\nu}\partial_\mu\phi\partial_\nu\phi+\alpha f(\phi)\mathcal{L}_{\mathrm{GB}}\bigg]}\\
    -\frac{1}{8\pi G_{N}}\int_{Q}{d^{2}x\sqrt{h}\bigg(K+\alpha f(\phi)\delta^{i_1i_2i_3}_{j_1j_2j_3}K^{j_1}_{i_1}\bigg(\overline{R}^{j_2j_3}_{\:\:\:\:\:\:\:\:i_2i_3}-\frac{2\epsilon}{3}K^{j_2}_{i_2}K^{j_3}_{i_3}\bigg)\bigg)}\label{eq:H.2}
    \end{gathered}
\end{equation}

In the AdS/CFT correspondence the IR divergences in the gravity side, correspond to the UV divergences at CFT boundary theory, which relation is known as the IR-UV connection. In our case for Lovelock theory, we have to the AdS-BTZ black hole, removing the IR divergence that:

\begin{equation}
     \begin{gathered}
    I_{bulk}=-\frac{1}{16\pi G_{N}}\int^{2\pi r_{h}}_{0}{d\tau}\int^{y}_{y_{0}}{dy}\int^{r_{h}}_{r_{b}}{\frac{Ldr}{r^{3}}\bigg[\kappa R-2L^{2}g^{rr}\phi^{'2}+\alpha L^{2}f(\phi)\mathcal{L}_{\mathrm{GB}}\bigg]}\\
    -\frac{1}{8\pi G_{N}}\int^{2\pi r_{h}}_{0}{d\tau}\int^{y}_{y_{0}}{{dy}\frac{\sqrt{f(r_{b})}}{r^{2}_{b}}\bigg(K_{(\Bar{\gamma})}+\alpha f(\phi)\delta^{i_1i_2i_3}_{j_1j_2j_3}K^{j_1}_{i_1 \:\:\:\:(\Bar{\gamma})}\bigg(\overline{R}^{j_2j_3}_{\:\:\:\:\:\:\:\:i_2i_3}-\frac{2\epsilon}{3}K^{j_2}_{i_2}K^{j_3}_{i_3\:\:\:\:(\Bar{\gamma})}\bigg)\bigg)}\label{eq:H.3}
    \end{gathered}
\end{equation}

Where $r_{b}$ is a cutoff, that has been to included to remove the divergence of IR regime. On the other hand, we have the following quantities

\begin{equation}
     \begin{gathered}
     \mathcal{L}_{\mathrm{GB}}=R^2-4R_{\mu\nu}R^{\mu\nu}+R_{\mu\nu\alpha\beta}R^{\mu\nu\alpha\beta}\\
     R=-[6f(r)-4rf^{'}(r)+4r^{2}f^{''}(r)]\\
    -4R_{\mu\nu}R^{\mu\nu}=\bigg(-\frac{[4f+r(-3f^{'}+rf^{''})]^{2}}{L^{4}},-\frac{4(-2f+rf^{'})^{2}}{L^{4}},-\frac{[4f+r(-3f^{'}+rf^{''})]^{2}}{L^{4}}\bigg)
    \end{gathered}
\end{equation}

Using the above quantities, we have

\begin{equation}
     \begin{gathered}
    I_{bulk}=-\frac{r_{h}L\Delta y}{8G_{N}}\int^{r_{h}}_{r_{b}}{\frac{dr}{r^{3}}\bigg[\kappa R-2L^{2}g^{rr}\phi^{'2}+\alpha L^{2}f(\phi)\mathcal{L}_{\mathrm{GB}}\bigg]}+\mathcal{O}(r_{b})\label{eq:H.4}
    \end{gathered}
\end{equation}

Now, considering the analogously step by step, we have for the boundary term that:

\begin{equation}
     \begin{gathered}
    I_{bdry}=-\frac{r_{h}L}{8G_{N}}\int^{r_{h}}_{r_{b}}{\frac{dr \Delta y(r)}{r^{3}}\bigg[\kappa R-2L^{2}g^{rr}\phi^{'2}+\alpha L^{2}f(\phi)\mathcal{L}_{\mathrm{GB}}\bigg]}\\
    -\frac{r_{h}L^{2}}{4G_{N}}\int^{r_{h}}_{r_{b}}{dr\frac{g(r)}{r^{2}}\bigg(K+\alpha f(\phi)\delta^{i_1i_2i_3}_{j_1j_2j_3}K^{j_1}_{i_1}\bigg(\overline{R}^{j_2j_3}_{\:\:\:\:\:\:\:\:i_2i_3}-\frac{2\epsilon}{3}K^{j_2}_{i_2}K^{j_3}_{i_3}\bigg)\bigg)}+\mathcal{O}(r_{b})\label{eq:H.5}
    \end{gathered}
\end{equation}
Here $K$ is given by:

\begin{equation}
     \begin{gathered}
    K=\frac{2(2f-rf^{'})y^{'}+f(4f-rf^{'})y^{'3}-2rfy^{''}}{2Lg^{3}(r)}
    \end{gathered}
\end{equation}

As we can see your results to the holographic renormalization would be available by the numerically procedure. Beyond, the Euclidean action is given by $I_{E}=I_{bulk}-2I_{bdry}$:

\begin{equation}
     \begin{gathered}
    I_{E}=-\frac{r_{h}L\Delta y}{8G_{N}}\int^{r_{h}}_{r_{b}}{\frac{dr}{r^{3}}\bigg[\kappa R-2L^{2}g^{rr}\phi^{'2}+\alpha L^{2}f(\phi)\mathcal{L}_{\mathrm{GB}}\bigg]}\\
    -\frac{r_{h}L}{4G_{N}}\int^{r_{h}}_{r_{b}}{\frac{dr \Delta y(r)}{r^{3}}\bigg[\kappa R-2L^{2}g^{rr}\phi^{'2}+\alpha L^{2}f(\phi)\mathcal{L}_{\mathrm{GB}}\bigg]}\\
    -\frac{r_{h}L^{2}}{2G_{N}}\int^{r_{h}}_{r_{b}}{dr\frac{g(r)}{r^{2}}\bigg(K+\alpha f(\phi)\delta^{i_1i_2i_3}_{j_1j_2j_3}K^{j_1}_{i_1}\bigg(\overline{R}^{j_2j_3}_{\:\:\:\:\:\:\:\:i_2i_3}-\frac{2\epsilon}{3}K^{j_2}_{i_2}K^{j_3}_{i_3}\bigg)\bigg)}\label{eq:H.6}
    \end{gathered}
\end{equation}

In the above equation, the second term in $I_{E}$ have the presence of the boundary profile $y(r)$, which characterize the BTZ black hole in Lovelock theory. An analytical solutions are possible in this gravity scenario, but we need impose special truncation \cite{PhysRevD.104.066014}. In fact, we want to analyze the problem in a numeric way without restriction inside the theory. In our prescription, we want to understand the BCFT$_{2}$ behavior for such gravity theories.

\section{BTZ thermodynamics in the Gauss-Bonnet gravity}\label{sec:7}
In this subsection, we are going to obtain thermodynamical quantities for AdS$_3$ BTZ black holes in the presence of manifold boundary. Firstly, we are going to start from black hole entropy. 
\subsection{BTZ entropy}

In order to compute the thermodynamics quantities, we need of the  entropy related to the BTZ black hole that has contributions of the AdS/BCFT correspondence in Lovelock theory. Thus, our the free energy defined as

\begin{equation}
     \begin{gathered}
     F=T_{H}I_{E}\label{eq:Free.1}
    \end{gathered}
\end{equation}
Using the above, we can compute the corresponding entropy as:

\begin{equation}
     \begin{gathered}
     S=-\frac{\partial F}{\partial T_{H}}=-\bigg(I_{E}+T_{H}\frac{\partial I_{E}}{\partial T_{H}}\bigg)_{numerically}\label{eq:Free.2}
    \end{gathered}
\end{equation}
Where
\begin{equation}
\begin{gathered}
    \frac{\partial I_E}{\partial T_H}=\frac{\partial I_E}{\partial r_h}\frac{\partial r_h}{\partial T_H}=\frac{2\pi r_h^2}{8G_N}\bigg(L \bigg(\Delta y \int_{r_b}^{r_h} \mathcal{I}_1(r) \, dr+2
   \int_{r_b}^{r_h} \mathcal{I}_2(r) \, dr\\
   +4 L
   \left(\int_{r_b}^{r_h} \mathcal{I}_3(r) \, dr+r_h
   \mathcal{I}_3(r_h)\right)+r_h \Delta y \mathcal{I}_1(r_h)+2 r_h
   \mathcal{I}_2(r_h)\bigg)\bigg)
\end{gathered}
\end{equation}
Here $\mathcal{I}_1$, $\mathcal{I}_2$ and $\mathcal{I}_3$ denote the integrands of Euclidean action (\ref{eq:H.6}).
The idea behind our dictionary is replace the numerical value of the temperature in $I_{E}$ and extract the entropy.
\begin{table}[!htbp]
\centering
\begin{tabular}{llllll}
\hline
$S$                   & $\alpha$ & $L$                    & $S$                   & $\alpha$ & $L$ \\ \hline
$1.048\times 10^{11}$ & 1        & \multicolumn{1}{l|}{1} & $3.000\times 10^{11}$ & 1        & 1.1 \\
$3.646\times 10^{10}$ & 1.1      & \multicolumn{1}{l|}{1} & $9.509\times 10^{10}$ & 1.1      & 1.1 \\
$1.491\times 10^{10}$ & 1.2      & \multicolumn{1}{l|}{1} & $3.620\times 10^{10}$ & 1.2      & 1.1 \\ \hline
\end{tabular}
\label{tab:1}
\caption{Black hole entropy values for different $\alpha$ and $L$, for unitary BTZ black hole horizon radii $r_h=1$}
\end{table}

As one may obviously notice from the Table (\ref{tab:1}) data, black hole entropy is getting smaller with $\alpha\to\infty$ and bigger with the grow of AdS radius $L$. This regime of parameters control the information storage, its adds limitations to the BTZ black hole entropy where this information is bounded by the BTZ black hole area. Thus, when an object such as a black hole captures mass it can be forced to undergo a gravitational collapse and the second law of thermodynamics insists that it must have less entropy than the resulting black hole, which for $\alpha\to\infty$ is very small, decreasing the entropy.

\subsection{Other thermodynamical quantities}
In order to analyze the stability of the BTZ black we compute the thermodynamics quantities, as for example,  heat capacity, sound speed and the trace of the energy-momentum tensor, all this relation comes from the canonical Ensemble. Such quantities following the prescription of \cite{PhysRevD.104.066014} are, respectively, given by

\begin{equation}
     \begin{gathered}
     C_{V}=T\bigg(\frac{\partial S}{\partial T}\bigg)_{V}=-T\bigg(\frac{\partial^{2} F}{\partial T^{2}}\bigg)\\
     c^{2}_{s}=\frac{S}{C_{V}}\\
      \langle T^a_{\ \ a}\rangle = 4F + TS\label{eq:QUANT.1}
    \end{gathered}
\end{equation}
As well, we could define
\begin{equation}
\begin{gathered}
    \frac{\partial S}{\partial T}=\frac{\partial S}{\partial r_h}\frac{\partial r_h}{\partial T}= \frac{1}{8 G_N}\bigg(L \bigg(\Delta y \int_{r_b}^{r_h} \mathcal{I}_1(r) \, dr+2 \int_{r_b}^{r_h} \mathcal{I}_2(r) \, dr+4 L
   \int_{r_b}^{r_h} \mathcal{I}_3(r) \, dr\\
   +r_h \mathcal{D}T \Delta y \mathcal{I}_1'(r_h)+\Delta y (r_h+2 \mathcal{D}T) \mathcal{I}_1(r_h)+2
   r_h \mathcal{D}T \mathcal{I}_2'(r_h)+2 (r_h+2 \mathcal{D}T) \mathcal{I}_2(r_h)\\
   +4 r_h \mathcal{D}T L \mathcal{I}_3'(r_h)+8 \mathcal{D}T
   L \mathcal{I}_3(r_h)+4 r_h L \mathcal{I}_3(r_h)\bigg)\bigg)
   \end{gathered}
\end{equation}
Here $F=T_{H}I_{E}$, $\mathcal{D}T=\partial r_h/\partial T$ and $C_{V}$ says as the BTZ black hole can be stable or unstable. On the other hand, the sound speed given by $c^{2}_{s}$, says as system can be has a behavior conformal. Such value of this behavior is around of $c^{2}_{s}\to 1/3$, we expect found these value numerically. However, the $\langle T^a_{\ \ a}\rangle/T^{4}\to 0$ for the high-temperature regime, its must be recovery the conformal symmetry and therefore the emergence of a nontrivial BCFT in Lovelock theory. 
\begin{table}[!htbp]
\centering
\begin{tabular}{llllll}
\hline
$C_V$                 & $\alpha$ & $L$                    & $C_V$                 & $\alpha$ & $L$ \\ \hline
$2.4583\times10^{15}$ & 1        & \multicolumn{1}{l|}{5} & $2.4573\times10^{15}$ & 1        & 5.1 \\
$2.7326\times10^{15}$ & 1.1      & \multicolumn{1}{l|}{5} & $2.5483\times10^{15}$ & 1.1      & 5.1 \\
$2.955\times10^{15}$  & 1.2      & \multicolumn{1}{l|}{5} & $2.9704\times10^{15}$ & 1.2      & 5.1 \\ \hline
\end{tabular}
\label{tab:2}
\caption{Numerically derived black hole heat capacity for BTZ black hole interior with $r_h=1$}
\end{table}

We computed black hole heat capacity for various values of $\alpha$ and $L$ and stored them in the Table (\ref{tab:2}). During the numerical analysis, we noticed that $C_V>0$ (stable solution) only for the case with $L\gg0$. Moreover, only this case correspond to the BTZ black hole within AdS/BCFT that respect causal conditions ($c_s^2\geq0$).

\begin{table}[!htbp]
\centering
\begin{tabular}{llllll}
\hline
$c_s^2$   & $\alpha$ & $L$                    & $c_s^2$   & $\alpha$ & $L$ \\ \hline
$0.25193$ & 1        & \multicolumn{1}{l|}{5} & $0.24213$ & 1        & 5.1 \\
$0.25197$ & 1.1      & \multicolumn{1}{l|}{5} & $0.24222$ & 1.1      & 5.1 \\
$0.25202$ & 1.2      & \multicolumn{1}{l|}{5} & $0.24222$ & 1.2      & 5.1 \\ \hline
\end{tabular}
\label{tab:3}
\caption{Black hole sound of speed w.r.t. Lovelock coupling constant $\alpha$ and AdS radius $L$ with $r_h=1$}
\end{table}

Besides, we as well calculate the sound of speed numerically from expression (\ref{eq:QUANT.1}), place the solutions in the Table (\ref{tab:3}). Obviously, one may notice that for $L\gg0$ causality conditions are validated, speed of sound squared does not exceeds speed of light. As well, here we see that $c_s^2$ is near $1/3$, so as expected system could behave like conformal one.

Finally, we numerically derive the trace of energy-momentum tensor for $T\to\infty$. We have noticed that for very small values of event horizon radii $r_h$ (that corresponds to the high temperature regime), $y(r)$ has constant (UV) solution only, other ones are complex. Therefore, as prescribed $\langle T^a_{\:\:\:\:a} \rangle/T^4$ vanish when $T\to\infty$, which signifies the arise of conformal symmetry.

As the last quantity, we present the Hawking-Page phase transition. For this, following the results of \cite{PhysRevD.104.066014}, we have that $\Delta \mathcal{V}_{E}=\mathcal{V}_{E}-\mathcal{V}_{E}(0)$ where

\begin{equation}
     \begin{gathered}
    \mathcal{V}_{E}=\frac{r_{h}L\Delta y}{8G_{N}}\int^{r_{h}}_{r_{b}}{\frac{dr}{r^{3}}\bigg[\kappa R-2L^{2}g^{rr}\phi^{'2}+\alpha L^{2}f(\phi)\mathcal{L}_{\mathrm{GB}}\bigg]}\\
    +\frac{r_{h}L}{4G_{N}}\int^{r_{h}}_{r_{b}}{\frac{dr \Delta y(r)}{r^{3}}\bigg[\kappa R-2L^{2}g^{rr}\phi^{'2}+\alpha L^{2}f(\phi)\mathcal{L}_{\mathrm{GB}}\bigg]}\\
    +\frac{r_{h}L^{2}}{2G_{N}}\int^{r_{h}}_{r_{b}}{dr\frac{g(r)}{r^{2}}\bigg(K+\alpha f(\phi)\delta^{i_1i_2i_3}_{j_1j_2j_3}K^{j_1}_{i_1}\bigg(\overline{R}^{j_2j_3}_{\:\:\:\:\:\:\:\:i_2i_3}-\frac{2\epsilon}{3}K^{j_2}_{i_2}K^{j_3}_{i_3}\bigg)\bigg)}\label{eq:H.6}
    \end{gathered}
\end{equation}

and $\mathcal{V}_{E}(0)$ corresponds to the thermal AdS BTZ black hole. In this case $f(r)\to 1$, providing 

\begin{equation}
     \begin{gathered}
     \mathcal{L}_{\mathrm{GB}}=R^2-4R_{\mu\nu}R^{\mu\nu}+R_{\mu\nu\alpha\beta}R^{\mu\nu\alpha\beta}\\
    R=-6,\quad-4R_{\mu\nu}R^{\mu\nu}=\bigg(-\frac{16}{L^{4}},-\frac{16}{L^{4}},-\frac{16}{L^{4}}\bigg)\\
        K=\frac{4y^{'}+4y^{'3}-2ry^{''}}{2Lg^{3}(r)},\quad g^{2}(r)=1-y^{'2}(r)
    \end{gathered}
\end{equation}
For thermal AdS black holes, solution for $y(r)$ is constant and equals to the UV cut $y_0$, so $\Delta y=0$. Therefore:
\begin{equation}
     \begin{gathered}
    \mathcal{V}_{E}(0)=
    \frac{r_{h}L^{2}}{2G_{N}}\int^{r_{h}}_{r_{b}}{dr\frac{g(r)}{r^{2}}\bigg(K+\alpha f(\phi)\delta^{i_1i_2i_3}_{j_1j_2j_3}K^{j_1}_{i_1}\bigg(\overline{R}^{j_2j_3}_{\:\:\:\:\:\:\:\:i_2i_3}-\frac{2\epsilon}{3}K^{j_2}_{i_2}K^{j_3}_{i_3}\bigg)\bigg)}
    \end{gathered}
\end{equation}
Remarkably, for constant $y(r)$ we have that integral above also vanish, since $K=0$ and $K_{\mu\nu}=0$, so $\mathcal{V}_{E}(0)=-I_E(0)=0$. Besides, the Euclidean action integrated up to the BH event horizon for AdS$_3$ black hole is negative (therefore partition function is positive), hence $\Delta \mathcal{V}_E>0$ and thermal AdS spacetime is stable.
\section{Concluding remarks}\label{sec:8}
In the current section we are going to mentioned the key results of our investigation. Throughout our study, we considered Lovelock theory with non-minimally coupled to gravitation sector canonical scalar field and probed the AdS/BCFT correspondence for three dimensional BTZ black hole with 2D CFT at $r=0$.

We derived metric function and scalar field, constructed the profile of additional boundary for higher dimensional (bulk) AdS spacetime by solving the Equations of Motion for bulk and aforementioned boundary with modified gravitation theory (therefore, usual Gibbons-Hawking-York surface term were also modified conveniently). It was shown that both metric function and boundary profiles behave as expected. For graphical representation and more detailed discussion, see Figures (\ref{fig:3}) and (\ref{fig:4}).

Moreover, to get rid of the IR (at $r\to\infty$) and UV (at $r\to0$) divergences, we used the procedure of the so-called holographic renormalization for Lovelock theory of second order. With the use of Euclidean action integral, obtained from the holographic renormalization method, we derived free energy and holographic dual of the black hole entropy numerically using Runge-Kutta solver of fourth order. As it turned out, black hole entropy had negative values, which is necessary for physical viability of the theory. As well, we noticed that black hole entropy decrease with $\alpha\to\infty$, where $\alpha$ is Gauss-Bonnet coupling and increase with AdS radius $L$. Numerical solutions for black hole entropy with various values of free parameters are stored in the Table (\ref{tab:1}).

From the entropy and free energy of BTZ black hole it was possible to derive other thermodynamical quantities, such as heat capacity, sound of speed squared, normalized trace of the energy-momentum tensor or Hawking-Page phase transition. From the comprehensive numerical analysis of heat capacity, we noticed that it's values are positive (which is important for black hole interior thermodynamical stability) only for sufficiently big AdS radius $L\gg0$. On the other hand, this statement as well converges with the speed of sound, which $c_s^2\geq0$ only if $L\gg0$. As a final note, we want to discuss the Hawking-Page phase transition. We found out that numerical solutions of partition function for thermal AdS vanish and therefore $\Delta \mathcal{V}_E=\mathcal{V}_E$. Moreover, since Euclidean integral for regular asymptotically AdS BTZ black hole has negative behavior, $\Delta \mathcal{V}_E>0$ and therefore thermal AdS spacetime is stable.

However, in the future works we think that it will be of special interest to probe AdS/BCFT correspondence for other viable cosmological theories of gravitation with or without exotic matter fields, such as inflaton field, vector and scalar bosons, gravitons or $q$-forms.

\end{document}